\documentstyle[aps,prl,array,epsf]{revtex}

\newcommand{\be}{\begin{equation}}
\newcommand{\ee}{\end{equation}}
\newcommand{\ba}{\begin{array}}
\newcommand{\ea}{\end{array}}

\begin{document}

\title{Dynamic exponent in Extremal models of 
pinning.}

\author{Supriya Krishnamurthy$^1$, Anne Tanguy$^2$ and St\'ephane Roux$^3$} 

\address{ 
1: Laboratoire de Physique et de
M\'ecanique des Milieux H\'et\'erog\`enes, \\
Ecole Sup\'erieure de Physique et Chimie Industrielles de Paris, \\
10, rue Vauquelin, 75231 Paris Cedex 05,  France.\\
2: Universit\'e de Lyon I, 43 Bd. du 11 Novembre 1918,\\
69622 Villeurbanne, France.\\
3: Laboratoire Surface du Verre et Interfaces,\\
Unit\'e Mixte de Recherche CNRS/Saint-Gobain,\\
39 Quai Lucien Lefranc, BP 135,
F-93303 Aubervilliers Cedex, France.}
\date{\today}
\maketitle

\begin{abstract} 
The depinning transition of a front moving in a time-independent random
potential is studied.  The temporal development of the overall roughness
$w(L,t)$ of an initially flat front, $w(t)\propto t^\beta$, is the
classical means to have access to the dynamic exponent.  However, in the
case of front propagation in quenched disorder via extremal
dynamics, we show that the initial increase in front roughness implies
an extra dependence over the system size which comes from the fact that
the activity is essentially localized in a narrow region of space.  We
propose an analytic expression for the $\beta$ exponent and confirm
this for different models (crack front propagation, 
Edwards-Wilkinson model in a quenched noise, ...).  
\end{abstract}




The propagation of a front in a noisy environment has been the subject of
active research in the past.  In particular, after Family and Vicsek
proposed a scaling form for the evolution of the roughness of fronts and
surfaces from Langevin equations, numerous analytical and numerical
works have verified these laws in a wide variety of models.  Various
reviews cover this rich field\cite{FamilyVicsek,HalpinZhang,Meakin}.

More recently, the quenched (i.e. time-independent) nature of the noise was
recognised as  playing a significant role in front propagation. 
Unfortunately, in spite of a few key works, analytic modelling of
such a depinning transition is rather scarce:  Dynamic Renormalization
Group studies have been proposed for the Edwards-Wilkinson model with 
quenched disorder\cite{Leschhorn1,Leschhorn2,Narayan} however in 1+1
dimensions some of the predicted exponents are quite far from their 
estimate obtained in numerical simulations. It has been proposed that  
the Kardar-Parisi-Zhang
model in a quenched environment can be described by a
directed percolation related model, mostly on the basis of numerical
agreement between measured exponents\cite{Buldyrev,Tang}.

Such models are relevant for a number of pinning phenomena such as
crack propagation \cite{Bouchaud,EB,SchmMaloy,Ramanathan,Gao}, wetting
phenomena \cite{Joanny,Paterson}, vortex pinning in type-II 
superconductors \cite{Larkin} and solid friction\cite{Pearson,Caroli}. 

Growth models in 1+1 dimensions are often studied through the evolution
in time $\theta$, (defined as the number of growth steps $t$ divided by
the system size $L$), of the roughness, $w(\theta)$, of an initially flat front
$w(0)=0$.  The roughness --- standard deviation of the front --- in a
system of size $L$ obeys the scaling form
\be\label{eq1}
w(\theta)=\theta^\beta\varphi\left({L\over \theta^{1/z_1}}\right)
\ee
where 
\be
\varphi(x)=\cases{cst & for $x\ll 1$\cr
x^\zeta & for $x\gg 1$}
\ee
and $\zeta=\beta z_1$.  The exponent $z_1$ is referred to as the dynamic
exponent since it relates space and time, whereas $\zeta$ describes the
roughness of the front (Hurst or roughness exponent), i.e. the scaling
of the pair correlation function for equal-time positions along front.  
In most cases of annealed noise, there is no need\cite{Lopez} 
to introduce any other
exponents, since at late stages, when the overall roughness has reached
the saturation value, the full two point correlation function of the front
at different locations and times reveals a similar scaling between space
and time: $\Delta x\propto \Delta \theta^{1/z_1}$.

The aim of this letter is to show that for a class of quenched disorder
depinning models, the early time development of the roughness does not
obey such a law. Rather, it implies an extra $L$ dependence in the
expression of $w$, from which $\zeta=\beta z_1$ is violated.

We focus more
specifically on a class of models introduced by Tanguy {\it et
al}\cite{Tanguy1} obeying extremal dynamics.  The front is defined
by its position $z=h(x,t)$.  An external driving $F$ allows to exert a
pressure on the front which is however biased by the position of the
front, so that at each site $x$, the force is $f(x,t)=F(G\star h)$ where
the $\star$ denotes a convolution product, and $G$ is a function which
is specific to the physical problem studied.  The local part of $G$ is
adjusted so that $G(0)= - \sum_{x\neq 0} G(x)$.   In order to normalize
this kernel, we choose as a convention $G(0)=-1$.  The key feature is that
$G$ may have power-law tails which determine the universality class of
the problem. For a planar crack, Gao and Rice \cite{Gao} have shown that (to first
order in $h$), $G$ decays as $r^{-2}$.  The same holds for the motion of
the triple line of a liquid-glass interface intersecting a solid surface,
in a wetting problem with a semi-infinite liquid surface in weak
gravity.  If the liquid/gas system is confined between
two parallel plates the $G$ function decays more abruptly, as $r^{-3}$.
In the mean-field limit, the $G$ function does not depend on distance.  
For the sake of convenience, this model has thus been generalized to any
power-law form for $G$:
\be
G(r)\propto r^{-\alpha}
\ee
In our simulations, periodic boundary conditions are implemented and thus
we adjust $G$ to match such a periodicity: namely $G(r)\propto
\sin(\pi r/L)^{-\alpha}$.  Such an expression is exact for $\alpha=2$,
and is a convenient means to implement the periodic b.c. in other cases.

The environment is represented by quenched heterogeneities, $\eta(x,z)$,
which can block the front whenever $f(x,t)<\eta(x,h(x,t))$.  The $\eta$
values are considered to be uncorrelated, positive numbers, randomly
picked from a uniform distribution over the interval $[0;1]$.  Thus for
$F=0$, the interface does not move.  At each time step, the force is
increased slowly from 0 up to the level where one site $x^*$ 
(the active site) can depin.
This site jumps to the next obstacle $\eta$ at a random distance along
the $z$ direction.  This distance is chosen again from a uniform
distribution over the interval $[0;d]$.  The external loading is
immediately brought back to zero so that no other sites can move
simultaneously.  The same step is repeated indefinitely.  $d$ is a
free parameter.  It has been checked that $d$ plays no role in the
statistical properties of the model in the steady state.

The steady state properties of this model have been studied in detail
numerically\cite{Schmittbuhl,Tanguy1,Tanguy2}.  It has been shown in
particular that for $\alpha\le 1$, the model is in the mean field
regime, as can be easily inferred from the $\alpha=0$ case. No spatial
structure appears, and the front is an uncorrelated white noise.  For
$\alpha\ge 3$, the long-range kernel is dominated by the short
wavelength cut-off, and is thus equivalent to the case 
where $G$ is the
second derivative of a Dirac distribution, i.e. the local force is
proportional to the external loading and the local curvature of the
front.  This is the Edwards-Wilkinson or ``Laplacian'' case with quenched
disorder. The front has a ``super-rough'' structure with a roughness
exponent $\zeta\approx 1.2$\cite{Hansen}.  The steady state properties
of this dynamics is fairly rich and a number of scaling properties can be
observed for the front structure, the time evolution of the activity,
and of the driving force necessary to depin the interface. In the
intermediate range $1<\alpha<3$, similar properties are observed with
scaling exponents which vary continuously with $\alpha$.  

One especially interesting property can be studied in order to
characterize the spreading of activity in space and time: Let $x^*(t_0)$ be
the active site at time $t_0$, and $x^*(t_0+\Delta t)$ at time $t_0+\Delta t$. Note
that here we use as a time the total number of moves $t$ rather than 
the number of moves per site $\theta=t/L$.  We
study the statistical distribution $p(\Delta x,\Delta t)$ of $\Delta
x=|x^*(t_0+\Delta t)-x^*(t_0)|$ for a fixed $\Delta t$.  It obeys the scaling form 
\be
p(\Delta x,\Delta t)=\Delta t^{-1/z_2}~
\Psi\left({\Delta x\over\Delta t^{1/z_2}}\right)
\ee
where 
\be
\Psi(x)=\cases{cst &for $x\ll 1$\cr
x^{-b} &for $x\gg 1$}
\ee
where $b$ is equal to $\alpha$.  Such a scaling form was first
introduced by Furuberg {\it et al} studying invasion percolation\cite{Furuberg}.
Thus, in the steady state, we see that the activity spreads typically over
distances $\Delta x\propto \Delta t^{1/z_2}$.  In this sense, $z_2$ is
actually the ``dynamic exponent'', and it indeed governs all correspondences
between time and space in the steady state.  Note however that a
different convention is now used for $z$, because of the definition of time.
In particular the activity has spread over the entire system for a time
equal to $t\propto L^{z_2}$, hence $\theta=t/L\propto L^{z_2-1}$.  Thus
actually $z_2$ should be compared to $z_1+1$.
In models with extremal dynamics, $z_2$ can easily be
related to the roughness exponent of the front\cite{Paczuski,Tanguy1}.  
After a time
$\Delta t$, the activity remains localized in a region of extend $\Delta
x$.  Over this region, the front moves by a typical distance of order 
$\Delta z\propto \Delta x^\zeta$.  Thus the number of time steps required to travel by
this amount scales as $\Delta x\Delta z\propto \Delta x^{1+\zeta}$,
hence
\be
z_2=1+\zeta
\ee
This simple argument has been checked to be obeyed precisely in
numerical simulations for $1<\alpha<3$.  Let us however note that it
breaks down in the Laplacian case $\alpha\to \infty$ where $z_2\approx 2.0$
and $\zeta\approx 1.25$.  However, the situation when $\zeta>1$ is
known to display some pathological behaviors.

\begin{figure}
\label{figa}
\centerline{\epsfxsize=0.7\hsize\epsffile{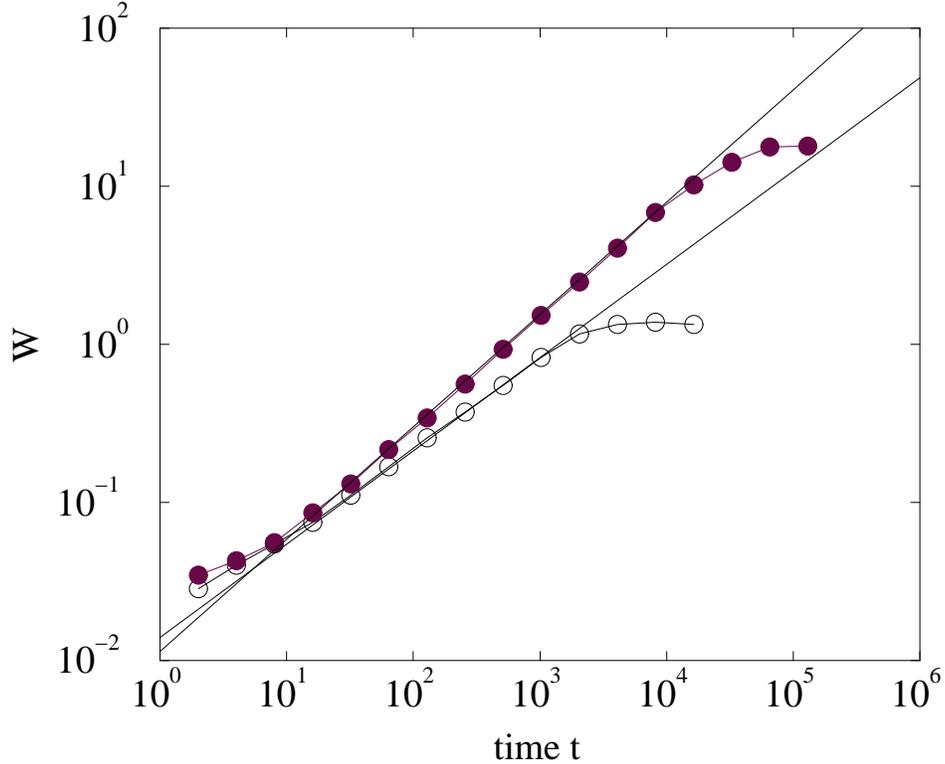}}
\caption{
Log-log plot of the overall roughness of the interface as a function of
time for two values of the $\alpha$ parameter $\alpha=2$ (symbol
$\circ$) and $\alpha=3$ (symbol $\bullet$).  The best power-law fits are
shown as plain lines.   The values of the slopes are reported in Table
\ref{taba}. The size of the system is L=1024.}
\end{figure}

We now consider the early stages of the front growth.  We assign an
initial flat configuration for the front $z=0$ and study the
evolution of the front roughness, in order to characterize $z_1$ through
the scaling relation Eq.~\ref{eq1}.  Figure ~1~\ref{figa} shows a log-log
plot of the overall roughness for the crack model, $\alpha=2$, and for 
$\alpha=3$.  In both cases, we indeed measure a power-law, $w\propto
t^\beta$, with $\beta\approx 0.61$ and 0.70 respectively.  
Table~\ref{taba} gives the
different values of $\zeta$ and $z_2$ from Ref.~\cite{Tanguy1}, and $\beta$
from the numerical simulations of the present study.  We note that if we
blindly apply the relation $z_1=\zeta/\beta$, we find 
a severe discrepancy with $z_2-1$ which
cannot be attributed to numerical uncertainties. For instance, if
$\alpha=2$, $\zeta/\beta=0.58$, and $z_2-1=0.35$.  Let us note that this
observation invalidates the numerical determination of the $z$ exponent
published for example in Refs.\cite{Leschhorn1,Leschhorn2} (moreover, in these articles the {\it analytical} expression proposed for $z$ , using RG analysis, refers to another driving mode that could be interesting 
to compare with extremal models).
This is the main message of this paper: The standard relation
$z=\zeta/\beta$ breaks down for quenched disorder growth models with
extremal dynamics.

\begin{figure}
\centerline{\epsfxsize=0.7\hsize\epsffile{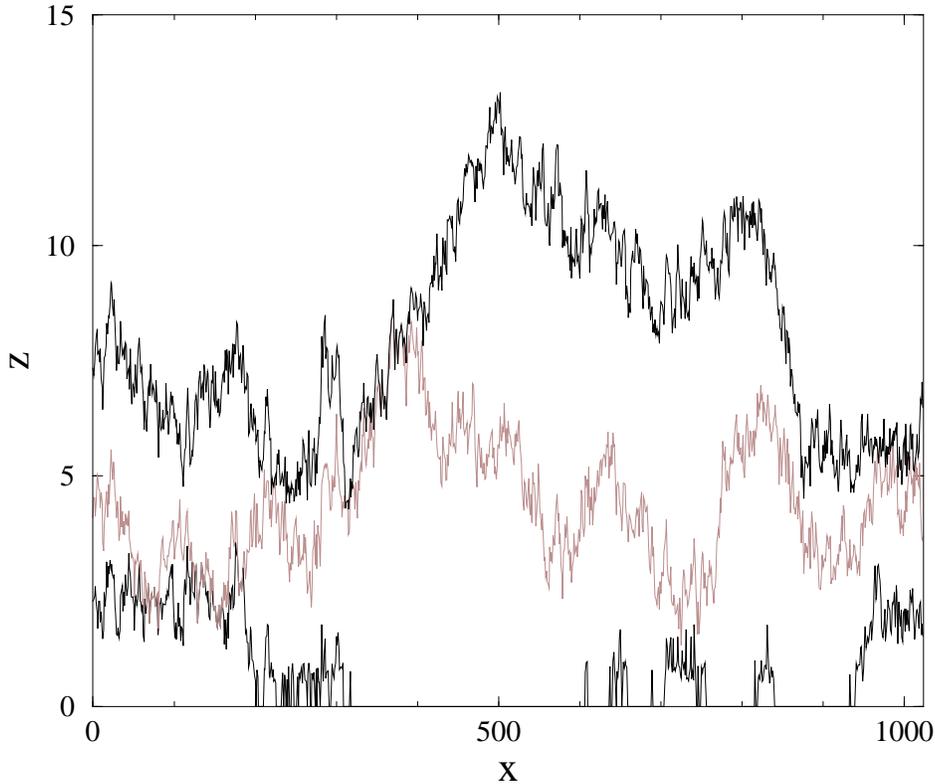}}
\label{figb}
\caption{
Shape of the first stages $t=1000$, 6000 and 11000 of the interface 
for $\alpha=2$ and $L=1024$. We note that the activity
is not equally distributed along the interface, but rather only a
localized part grows and reaches a steady state conformation before
growing and invading the flat region.  This observation expressed in
quantitative terms gives our prediction of $\beta$.}
\end{figure}

We can go one step beyond the numerical study reported so far and
estimate the value of $z_1$ from the steady state roughness exponent.
To this end, it is informative to look at the shape of the front in the
initial regime.  Figure ~2~\ref{figb} shows such an example for $\alpha=2$
and $L=1024$.  We observe
that the front remains pinned along its initial flat geometry for a
large time, and thus the interface moves only in a confined region of space. 
This region progressively grows, and invades the flat part, but still
continues its propagation along the $z$ direction.  This is in marked
contrast with annealed noise type growth where the activity is
delocalized along the entire front.  Moreover, as can be noted from
Figure ~1~\ref{figa}, the cross-over to the saturation regime is quite
steep.  This suggests that the depinned parts of the front have already
reached their steady state roughness.

Let us now translate this argument into quantitative terms.  We assume
that the interface is depinned along one single interval of length
$\ell$.  Along this interval, the roughness assumes its asymptotic
value, i.e. the typical height will be of order $h\approx \ell^\zeta$.
Therefore the time needed to reach this position is 
\be
t\propto \ell^{1+\zeta}
\ee
Generally, as one follows the time development of the roughness, one
introduces as a time the number of moves {\it per site}, $\theta=t/L$.
Ignoring subdominant terms coming from the average height of the
interface, we compute the overall roughness, $w$, as  
\be
w^2\propto {\ell^{2\zeta+1}\over L}
\ee
where the multiplicative $\ell/L$ term comes from the weight of the
depinned part of the interface as compared to the rest.  Therefore the 
$\beta$ exponent is readily estimated from the elimination of $\ell$ in
the two above equations, and thus  
\be
w\propto \theta^{\beta}L^{\beta-1/2}
\ee
with
\be\label{eqbeta}
\beta={\zeta+1/2\over\zeta+1}
\ee
We note however that there is an extra term depending on $L$ in the rate of
growth of the roughness.  Thus one cannot use the standard relation
$z_1=\zeta/\beta$.  Instead, we write
\be
w=\theta^{\beta}L^{\beta-1/2}\varphi\left({L\over \theta^{1/z_1}}\right)
\ee
The dynamic exponent $z_1$ is then found by imposing that in the steady
state $w$ is $t$-independent, and $w\propto L^\zeta$.  This gives the
power-law behavior of $\varphi$ for small arguments, $\varphi(x)\propto
x^{\zeta+1/2-\beta}$, and thus 
\be
\beta-{\zeta+1/2-\beta\over z_1}=0
\ee
or using our above expression for $\beta$
\be
z_1={\zeta+1/2-\beta\over\beta}
=\zeta
\ee 
Thus the total number of moves necessary to reach the steady state
scales as $t=\theta L\propto L^{1+\zeta}$, and thus we recover our
previous expression for $z_2=1+z_1$ where the difference of one simply
comes from the definition of time.
\begin{table}
\begin{center}
\begin{tabular}{|c|c|c|c|c|c|}
\hline
$\alpha$&$\zeta$&$z_2$&$\beta$&$\beta$\\
&from Ref.~\cite{Tanguy1}~&from Ref.~\cite{Tanguy1}~&(measured)&from Eq.~(\ref{eqbeta})\\
\hline\hline
1.5&0.05&1.05&0.49&0.52\\
2.0&0.35&1.35&0.61&0.63\\
2.5&0.65&1.65&0.65&0.70\\
3.0&1.0&2.0&0.70&0.75\\
$\infty$&1.25&2.0&0.8$^*$&0.78\\
\hline
\end{tabular}
\end{center}
\caption{Values of the exponents $\zeta$ and $z_2$ taken from
Ref.[21],
measured value of $\beta$ from the present work, and
predicted value from Eq.~(10). $^*$ refers to Ref.[23].}
\label{taba}
\end{table}


In Table~\ref{taba}, we have reported the value of the $\beta$ exponent
estimated from numerical simulation data, and the one obtained from
Eq.~\ref{eqbeta}.  We note an excellent numerical agreement for all
values of $\alpha$ studied.

More generally if other moments of the interface height are computed, a
different scaling is expected.  Indeed for the moment of order $m$, a
similar computation gives $\langle(h-\langle
h\rangle)^m\rangle^{1/m}\propto t^{\beta_m}$ with 
\be
\beta_m={\zeta+1/m\over\zeta+1}
\ee
Hence, in contrast with annealed models, different moments give rise to
different estimates of the $\beta$ exponent.  This again
underlines the fact that care has to be taken with the
interpretation of the latter exponent.

We mentioned that the part of the interface which has moved had already
a conformation representative of the steady state regime.  In order to
check this, we can perform the following test.  After the interface has
reached the steady state, we choose an arbitrary time $t_0$ and record
the position of the interface $h_0(x)=h(x,t_0)$.  Then as the simulation
continues, we study the incremental motion of the front, $\Delta h(x,t)=
h(x,t)-h_0(x)$.  The time evolution of the roughness of $\Delta h$
follows indeed the same law as the early stage $t^\beta$.  This shows
that the initial flat configuration behaves as any late stage
configuration.

\begin{figure}
\centerline{\epsfxsize=0.7\hsize\epsffile{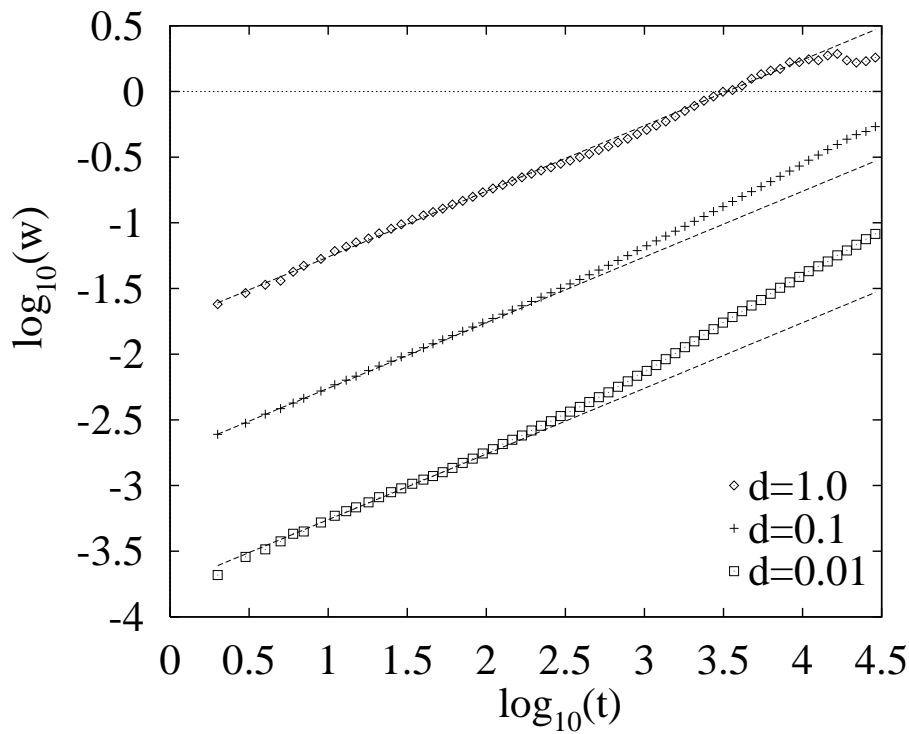}}
\caption{
Log-log plot of the overall roughness versus time for $\alpha=2$ and
$L=1024$.  Three values of $d$ are used as indicated in the caption.
The dotted line shows a slope of 0.5 which accounts for the very early
stages of growth when $d$ is small.  
}
\label{figc}
\end{figure}

The final question to answer is why the activity is localized during the
early stages of front propagation.  As a site moves, a large part of
the forces it carries is transferred to the nearest neighbours, while a
smaller part is transferred to the second neighbours, and so on.  The
amount of load transfer depends on the distance to the active site as
dictated by the kernel $G$ of the model, and hence it depends on
$\alpha$.  For $\alpha\to\infty$, the load transfer is local, and only
the nearest neighbors are influenced.  The amplitude of the change in
force is by the definition of the model dependent on the distance the
active site advances, and thus it depends on the parameter $d$.  For
large $d$, one may easily understand that the activity has a tendency to
move uniformly in the transverse direction.  
The first active site will jump by a distance
proportional to $d$, and hence its closest neighbours will be pushed
forward by such an amplitude that the amplitude of the threshold
strength may be insufficient to keep them pinned. In contrast, for small
$d$, the threshold distribution may win over the force modification
induced by the roughening of the front.  As the amplitude of $d$ is
reduced, indeed, the activity map shows that the very initial stage is
spread over the entire interface. In the very early stages, one measures
a roughness exponent $\beta\approx 0.5$ as can be expected from the
trivial observation that only a number $t$ of sites move by a single
step proportional to $d$, thus $w\propto d t^{1/2} L^{-1/2}$.  After
this initial transient, a higher slope takes over.
Figure ~3~\ref{figc} indicates  the evolution of $w(t)$ 
for three values of $d$, 1.0, 0.1
and 0.01. Fitting the time region where on average there has been more
than one move per site, $t>L$, gives consistent estimates of $\beta$ as
mentioned in Table~\ref{taba}.  We thus conclude that the above description
of the early stage growth is the generic case, which is encountered for any
specific choice of the parameter $d$.

Let us conclude by briefly summarizing our results: The time development
of roughness in quenched disorder depinning models implies a size
dependence which has not been noted before.  This implies a violation of
the scaling relation $\zeta=z\beta$.  Instead, we show that the activity
is localized even in the early stages of growth and thus this implies a
power-law increase of the roughness in time with an exponent $\beta$
given in Eq.~(\ref{eqbeta}).  This has been confirmed through numerical
simulations. 

\acknowledgments

We acknowledge the hospitality of the International Center of
Theoretical Physics in Trieste (It.) where this work was completed.



\end{document}